\documentclass[10pt,conference]{IEEEtran}

\usepackage{amssymb,amsmath} %,palatino}
\usepackage{multirow}
\ifCLASSINFOpdf
  \usepackage[pdftex]{graphicx}
\else
  \usepackage[dvips]{graphicx}
\fi
\usepackage{url}

\begin{document}

\title{On fuzzy syndrome hashing with LDPC coding}
\author{
  \IEEEauthorblockN{Marco Baldi, Marco Bianchi and Franco Chiaraluce}
  \IEEEauthorblockA{Universit\`a Politecnica delle Marche\\ 
  	Ancona, Italy \\
    Email: \{m.baldi, m.bianchi, f.chiaraluce\}@univpm.it
    }
\and
  \IEEEauthorblockN{Joachim Rosenthal and Davide Schipani}
  \IEEEauthorblockA{University of Zurich\\
    Zurich, Switzerland\\
    Email: \{rosenthal, davide.schipani\}@math.uzh.ch
}}

\maketitle

\begin{abstract}
The last decades have seen a growing interest in hash functions that allow some sort of tolerance, e.g. for the purpose of biometric authentication.
Among these, the syndrome fuzzy hashing construction allows to securely store biometric data and to perform user authentication without the need
of sharing any secret key.
This paper analyzes such a model, showing that it offers a suitable protection against information leakage and
several advantages with respect to similar solutions, like the fuzzy commitment scheme.
The design and characterization of LDPC codes to be used for this purpose is also addressed. 
\end{abstract}
%
%\vspace{2mm}
%\noindent
%{\bf Keywords:} biometric, fuzzy, hash function, information leakage
%
%\vspace{2mm}
%\noindent {\bf Mathematics Subject Classification (2010): } 94A60, 94A62, 94B05

% ********************************************************************

% ********************************************************************
%\vspace{8mm}

%\category{K.6.5}{Security and Protection}{Authentication}
%\category{E.4}{Coding and Information Theory}{Error control codes}


%\keywords{ACM proceedings, \LaTeX, text tagging}

\section{Introduction}

The use of biometric passwords, such as fingerprints, irises, etc., has been an important issue in recent years, both because of the big advantages it may bring along and because of the clearly non negligible privacy concerns and implementation issues \cite{ul04}.
 In fact, as far as privacy is concerned, the storage of raw biometric data is not an acceptable solution, but, 
on the other hand, a secure storage cannot be easily implemented, as for traditional passwords, by simply introducing an hash function. This is due to the fact that the binary strings derived from different acquisitions of the same biometric feature can slightly change from each other, and the biometric feature can slightly change itself. Therefore a certain threshold of tolerance is needed to be able to identify legitimate from non legitimate users, but this prevents the standard use of collision resistant hash functions \cite{sc10}. 

This problem has prompted researchers to devise other solutions for the secure storage and use of biometric passwords
(see \cite{tu07} for a selected survey of the literature).
%\cite{bo05,do04,fr01c,ha05b,ju06,ju99,sc10,tu07,ul04} 
%and the references therein, we limit ourselves to point out some of the most significant milestones. 
The idea behind most of these methods is a combined use of error correcting codes and hash functions, whose model is the fuzzy commitment scheme \cite{ju99}, which we revisit below. This has been later generalized to other types of metrics, such as the set difference metric \cite{ju06} and the edit distance metric \cite{do08}.

In particular, the fuzzy vault \cite{ju06} uses polynomial interpolation in order to allow authentication based on the matching of a sufficient number of features, while the fuzzy extractor \cite{do08} is a further generalization which combines the previous constructions with particular objects called random extractors. These make the previous schemes stronger with respect to information leakage, though they cannot prevent it \cite{bu08,do05}.

Briefly, privacy and implementation issues are still a concern and our aim is to give a further contribution, by showing that a syndrome based fuzzy hashing is actually feasible and even more convenient as far as information leakage is concerned.
First of all, we address the design of codes to be used within this context, with focus on LDPC codes.
% and provide both theoretical and numerical results concerning their design.
Secondly, we study the entropy of the fuzzy hashing output vectors and compare it with 
the corresponding quantity of its input vectors.

The paper is structured as follows: in Section 2 we briefly review the %prototype of these methods, that is the 
fuzzy commitment scheme and its main issues. Section 3 is devoted to the syndrome based construction which we denote
by fuzzy hashing. Section 4 addresses the design of codes to be used in this scheme. %, through both theoretical and numerical tools. 
Section 5 is aimed at analyzing the entropy of the output vectors, in order to assess the
performance of fuzzy hashing in terms of privacy.

\section{Prior art: the prototype}

%We briefly review here the prototype of most error correcting based methods, that is 
The fuzzy commitment scheme, proposed in \cite{ju99}, works as follows.
Suppose we want to securely store a length $n$ biometric vector $x \in \mathbb{F}_q^n$, where $\mathbb{F}_q$ is the Galois field of order $q$, and let $e$ be the maximum number of different 
symbols with respect to the reference vector $x$ that we can tolerate in any other acquisition of the same biometric feature.

According to the fuzzy commitment scheme, we choose a hash function $H_a$ and an $[n,k]$-linear block code $C \subset\mathbb{F}_q^n$,
able to correct $e$ errors, and then store $(H_a(r_x),l)$, where $r_x$ is a random codeword associated to $x$ and $l=x-r_x$.

Given another biometric $y$, we compute the vector $z = y-l$ and apply the decoding algorithm of $C$.
If decoding succeeds, this results in a codeword $c_z \in C$, and we compute $H_a(c_z)$.
If $H_a(c_z)$ equals $H_a(r_x)$, i.e., the value previously stored, we grant access, otherwise we deny it.

In fact, if the hashes are the same, then $c_z = r_x$ (apart from a negligible probability of a hash collision),
so $d(c_z, z) = d(r_x, z) \leq e$, where $d(\cdot)$ denotes the Hamming distance. 
Since $r_x = x-l$ and $z = y-l$, it results $d(r_x, z) = d(x, y) \leq e$.

Conversely, $d(x,y) \leq e$ implies $d(x-l,y-l) = d(r_x,z) \leq e$, so that decoding $z$ results in
$c_z = r_x$ and $H_a(c_z) = H_a(r_x)$.

In \cite{sc10}, we pointed out some of the main problems concerning the use of this scheme, namely implementation issues and security issues. In particular, privacy concerns may arise if the biometric templates are not uniformly distributed in the ambient space, that is their entropy is not maximal. In that case, it may be feasible to infer from $l$ some information about $r_x$ and, therefore, endanger the system security.

\section{Syndrome fuzzy hashing}

Starting from the fuzzy commitment principle, an alternative scheme (here fuzzy hashing) can be devised, in which syndromes
are used in the place of codewords.
Under a coding theory viewpoint, the two schemes are equivalent. Despite this, the use of syndromes has several advantages in 
the considered context.
The idea of storing the syndrome of $x$, instead of a shift vector from a codeword, already appeared in \cite{do08}, where it is
considered as an example of a sketch construction.
We will show that the use of fuzzy hashing is advantageous with respect to the classical fuzzy commitment scheme,
also by considering the characteristics of typical biometric data.

%First of all, although the relation with the fuzzy commitment scheme is very tight, this construction entails a better security profile, 
%if the entropy of the biometric templates is not maximal, and is therefore preferable.
%Secondly, for the codes needed in this context, the use of syndromes results in a considerable reduction of the stored data size,
%that is another important advantage when large data bases of biometrics are needed.

In the fuzzy hashing scheme, an $[n,k]$-linear block code $C \subset\mathbb{F}_q^n$, able to correct $e$ errors, is selected,
and it is described through its $r \times n$ parity-check matrix $H$, with $r = n-k$.
Given a biometric vector $x$ to be stored, the pair $(H_a(x),Hx)$ is used to represent $x$, were $H_a$ is a given hash function.
When another biometric $y$ is acquired and is compared with $x$, the value $Hx-Hy=H(x-y)=Hv$ is computed, that coincides with the syndrome associated to the difference vector $v = x-y$.
Then, syndrome decoding is applied on $Hv$, according to the chosen code $C$.
If $y$ is taken from the same individual as $x$, then $v$ has Hamming weight equal to $d(x,y) \le e$ and it corresponds
to a correctable error vector. So, syndrome decoding succeeds and correctly results in $v$.
Then, starting from $v$ and $y$, $x$ can be computed, as well as $H_a(x)$.
The latter coincides with the stored value, so access is granted.
Otherwise, syndrome decoding fails or reports $w \ne v$. In such case, $x' = w+y \ne x$
and $H_a(x') \ne H_a(x)$ is obtained, so access is denied.

In the fuzzy commitment, the vector $l=x-r_x$ is stored.
As some bits of the biometric $x$ might be known
with high probability, this reveals some information
on the secret codeword $r_x$.
The same may occur in fuzzy hashing, where the syndrome 
$Hx$ is stored, but only under the condition $x = q + r_q$,
where $q$ is a correctable error vector and $r_q$ is
any codeword.
In this case, syndrome decoding results in $q$; so,
some bits of $r_q$ can still be guessed, starting from
the predictable bits of $x$.
However, especially for very low rate codes, the probability that
$x$ is within the decoding radius of a codeword $r_q$
is very low, so fuzzy hashing provides better security
with respect to the classical fuzzy commitment.

\section{Codes for fuzzy hashing}

In order to design suitable codes to be included in the fuzzy hashing
scheme, we must consider the features of the biometric vectors we work with.
If we refer to fingerprints or irises, a common acquisition will consist of a vector
of several thousands of bits.
However, it would be unpractical to apply fuzzy hashing directly on the plain acquisition,
since a number of impairments could jeopardize the identification process.
In fact, small changes in the acquisition conditions (as ambient light or small movements
of the subject) could result in significant differences between two images of the same
biometric feature.
So, a common procedure is to extract a set of representative features from the
biometric data through algorithms aimed at making them invariant to some frequent 
acquisition impairments. An example of this kind of algorithms will be considered
in Section \ref{sec:EntropyAnalysys}.
So, the code must be designed to work with
the vectors produced as output by the feature extracting algorithm.
Typically, such vectors have length of the order of $10$k bits.

Another important aspect is the modeling of the errors affecting two vectors
resulting from different biometric acquisitions from the same individual.
In \cite{Sutcu2008}, the errors are modeled through a Binary Symmetric Channel (BSC)
with transition probability $p$.
%We will adopt this approach too.
We will adopt the same approach in this paper.
%In order to estimate the value of $p$ that is best suited to model this application,
%we can assume that typical feature extraction algorithms are used in the process.
As it will be shown in the next section, typical values of the percentage of different
bits between the vectors representing two acquisitions of the same biometric range
between $10$\% and $30$\%. So, we need codes with length about $10$k bits that are
able to correct such high fractions of errors and, hence, have very low rate ($R$).
Just to give an idea, a BCH code with ($n=2047$, $k=100$), that is, rate $R \approx 0.05$,
is able to correct $379$ errors, which means it has a relative error correcting capability of about $19$\%.
A BCH code with ($n=4095$, $k=110$), that is, rate $R \approx 0.03$, is able to correct $767$
errors, that is almost the same percentage.
A similar value is reached by the BCH code having ($n=8191$, $k=170$), hence rate $R \approx 0.02$,
able to correct $1533$ errors.

This evidences that, for classical algebraic codes, in order to maintain a given
relative error correcting capability, the code rate must be decreased as the code length
increases.
Furthermore, due to the long code length, 
decoding may also yield complexity issues, although recent algorithms can reduce the
decoding complexity \cite{Schipani2011}.
A smarter choice is represented by modern iteratively decoded error correcting codes,
like Low-Density Parity-Check (LDPC) codes \cite{Richardson2001}.
Actually, the use of LDPC codes in this context has already been proposed
in \cite{sc10, Sutcu2008, Bringer2007}, but the code design was not addressed
in those works.
In summary, fuzzy hashing with LDPC codes brings the following advantages:
\begin{itemize}
\item Fuzzy hashing reduces the amount of stored data,
with respect to the fuzzy commitment, since $r < n$.
%\vspace{-0.4em}
\item Fuzzy hashing reduces the predictability of the stored strings, 
as shown at the end of the previous section.
%\vspace{-0.4em}
\item LDPC codes have greater error correction
capabilities than classical algebraic codes. Moreover, their
relative error correcting capability, for a fixed rate,
is almost constant as the code length increases.
% This is also evidenced through density evolution, as it will be explained in the next subsection.
%\vspace{-0.4em}
\item LDPC codes allow to reduce the size of the code
representation, by exploiting the sparse nature of $H$.
\end{itemize}

\subsection{Code Design}

%In this subsection we aim at designing the codes needed in fuzzy hashing
%schemes, and at giving some possible choices of the code parameters for
%an actual implementation.

We are interested in almost regular codes, since they allow an easier implementation with
respect to irregular codes;
so, we fix the column weight of the matrix $H$ to be equal to an integer $d_v$.
The row weight, for the code rate values here of interest, cannot be constant
as well. However, it will be minimally dispersed around its mean 
$\left\langle d_c \right\rangle = d_v/(1-R)$.
If we suppose that (as it occurs for all the codes we consider):
\begin{equation}
\frac{k}{n} = R < \frac{1}{d_v+1},
\end{equation}
the matrix $H$ can have rows with only the following two values of weight: $d_v$ and $d_v+1$.
In this case, $r - k \cdot d_v$ rows have weight $d_v$ and the other 
$k \cdot d_v$ rows have weight $d_v + 1$.

We can describe the column and row weight distributions of the matrix $H$
through the polynomials $\lambda(x)$ and $\rho(x)$ representing, respectively,
the variable node and check node degree distributions of the associated Tanner graph \cite{Richardson2001}.
Since we adopt the edge perspective, $\lambda_i$ ($\rho_i$) denotes the fraction 
of ones in the parity-check matrix $H$ which are in columns (rows) of weight $i$.
Based on the hypotheses above, for the considered ensemble of codes it results:
\begin{align}
\lambda(x) & = x^{d_v-1}, \nonumber \\
\rho(x) & = [1 - R(1+d_v)]x^{d_v-1} + R(1+d_v)x^{d_v}.
\label{eq:lambda_rho}
\end{align}
Starting from \eqref{eq:lambda_rho},
we can estimate the asymptotic performance (that is, for $n \rightarrow \infty$) of LDPC
codes in this ensemble by applying the density evolution
method \cite{Richardson2001}.

Gallager's A algorithm \cite{Gallager1963} is an LDPC decoding algorithm for the BSC channel
that permits an easy characterization through density evolution \cite{Bazzi2004}.
So, we have estimated its convergence threshold (that is the maximum channel error probability such 
that all the errors can be corrected using an infinite length code) for the variable and check 
node degree polynomials given by \eqref{eq:lambda_rho}. Results are reported in Table \ref{tab:Thresholds},
where the threshold values computed for $d_v = 3, 4, 5$ are provided, for code rates
ranging between $0.1$ and $0.01$.

\begin{table}[htb]
\renewcommand{\arraystretch}{1}
\caption{Threshold values for the considered LDPC codes ensembles under Gallager's A decoding.}
\label{tab:Thresholds}
\centering
\scriptsize\begin{tabular}{|c||c|c|c|}
\hline
%\multirow{2}{*}{$R$} & \multicolumn{3}{|c|}{$d_v$} \\
%& 3 & 4 & 5 \\
$R$ & $d_v=3$ & $d_v=4$ & $d_v=5$ \\
\hline
\hline
0.1 & 0.159 & 0.078 & 0.045 \\
\hline
0.09 & 0.163 & 0.079 & 0.045 \\
\hline
0.08 & 0.166 & 0.08 & 0.046 \\
\hline
0.07 & 0.169 & 0.081 & 0.046 \\
\hline
0.06 & 0.173 & 0.083 & 0.047 \\
\hline
0.05 & 0.177 & 0.084 & 0.048 \\
\hline
0.04 & 0.18 & 0.085 & 0.048 \\
\hline
0.03 & 0.184 & 0.087 & 0.049 \\
\hline
0.02 & 0.188 & 0.088 & 0.05 \\
\hline
0.01 & 0.192 & 0.089 & 0.051 \\
\hline
\end{tabular}
\end{table}

As we observe from the table, the choice of a small value of $d_v$ (like $3$)
should be preferred.
On the other hand, the asymptotic performance under Gallager's A decoding
is not very good.
For example, to reach a relative error correcting capability of $19$\%,
for $d_v=3$, a code rate smaller than $0.02$ is required, that is
similar to that needed by a BCH code with $n=8191$.

Gallager's A algorithm allows an easy density evolution
analysis, that is useful to verify that LDPC codes can asymptotically reach 
the error correcting performance we need and for which values of code rate.
On the other hand,
when dealing with finite length codes, decoding algorithms with better
performance can be used.
In fact, Gallager's A algorithm is a majority-based algorithm exploiting
a fixed decision threshold $b$, that is not the most effective choice.
For example, adopting a variable $b$ (as in Gallager's B algorithm) gives a first 
performance improvement.
Furthermore, several improved versions of these algorithms have been proposed 
in the literature \cite{Miladinovic2005}, that are able to outperform Gallager's original algorithms.
%\cite{Liu2005}, \cite{Shan2005}.
Finally, the classical Sum Product Algorithm (SPA) \cite{Hagenauer1996}
%, based on the Belief Propagation principle, 
can also be applied on the BSC, even though,
in absence of soft-information from the channel, the initial likelihood associated
to each bit can assume only two opposite values.
Despite this, the SPA is able to significantly improve the error correction 
performance with respect to that predicted in Table \ref{tab:Thresholds}, 
as we will show in the next section, by providing some examples of practical codes.

\subsection{Examples}

In this subsection we provide examples of LDPC codes having
parameters of interest in the fuzzy hashing context.
The codes have been designed through the Progressive Edge 
Growth (PEG) algorithm \cite{Hu2001PEG}, by imposing almost constant
column and row weights for their parity-check matrices.
For given column and row weights, the PEG algorithm allows to design
finite length LDPC codes with very good performance under belief
propagation decoding.

In detail, we first fix the column weight $d_v$. Then, we impose
the lower triangular form for the parity-check matrices, in such a
way as to facilitate encoding, especially for very long codes.
This introduces a last column having weight $1$, and some columns
having weight $< d_v$. However, their incidence with respect to
the total number of columns is very small.
Then, the PEG algorithm is used to optimize the length of the local cycles
within the Tanner graph associated to each code, while
keeping the row weight distribution as much concentrated as possible.

So, the characteristics of the codes we have designed are well overlaid
with those fixed in the previous subsection.
The codes mentioned above have been used to perform Montecarlo simulations
over the BSC, based on SPA decoding.

\begin{figure}[htb]
\begin{centering}
\includegraphics[width=60mm,keepaspectratio]{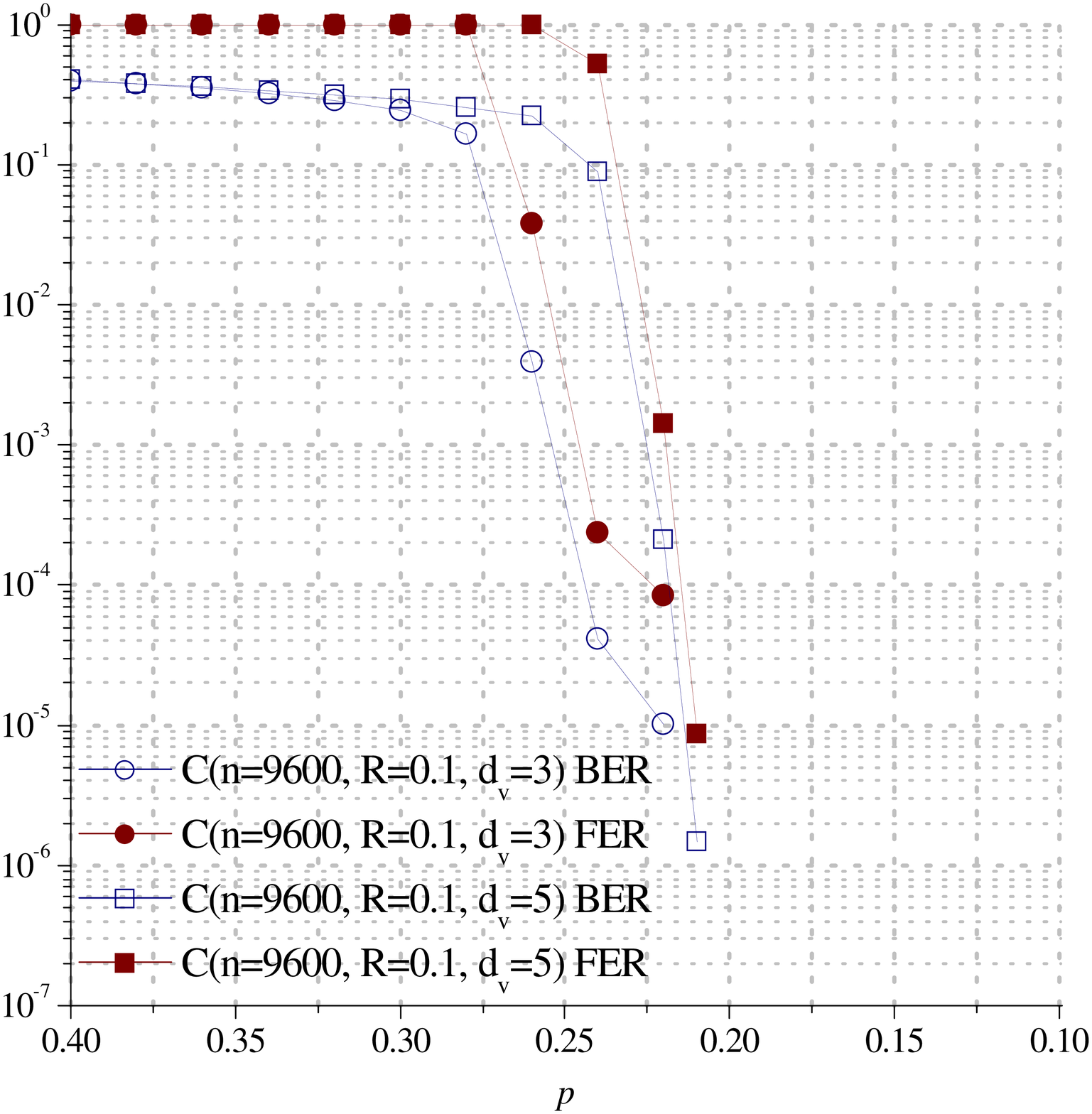}
\caption{Performance of rate $0.1$ LDPC codes with $d_v = 3$ and $d_v = 5$
over the BSC with SPA decoding.
\label{fig:Rate01}}
\par\end{centering}
\end{figure}

A first set of results is reported in Fig. \ref{fig:Rate01}, where LDPC
codes having $n=9600$ and $k=1000$ (hence, rate $\approx 0.1$) have been
considered. We have designed two codes with different column weights:
$d_v=3$ and $d_v=5$.
As we observe from the figure, the simulation confirms that the code with 
$d_v=3$ has better performance, in the waterfall region, with respect
to the code having $d_v=5$. This was expected on the basis of the results of
density evolution. However, we also observe that the code with $d_v=5$ has
a better performance in the error floor region, so its Bit Error Rate (BER)
and Frame Error Rate (FER) curves tend to intersect with those of the first
code. So, the choice of $d_v=3$ is suitable if a failure rate on the
order of $10^{-4}$ or more is acceptable; otherwise, the choice of $d_v=5$
should be preferred.

The performance improvement due to the SPA is
evident: both codes are able to achieve a rather low error rate for a
percentage of bit errors around $20$\%, or even more.

\begin{figure}[htb]
\begin{centering}
\includegraphics[width=60mm,keepaspectratio]{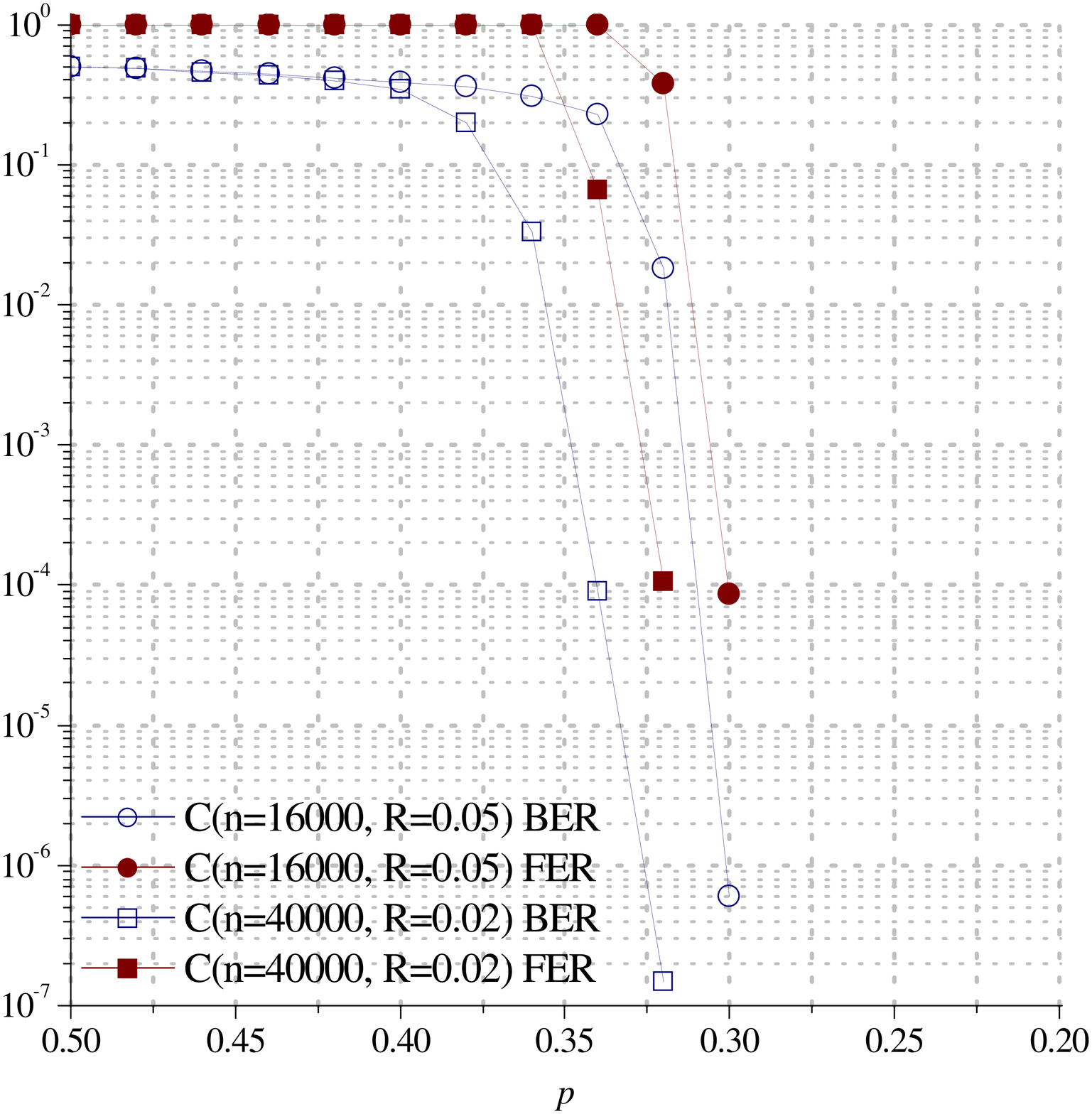}
\caption{Performance of LDPC codes with $d_v = 3$ and rate $0.05$ and $0.02$
over the BSC with SPA decoding.
\label{fig:OtherRates}}
\par\end{centering}
\end{figure}

To further increase the error correcting capability of these codes,
it is necessary to reduce their rate.
To provide some examples in this sense, we have considered $k=800$ and designed two
other LDPC codes, having $d_v=3$ and rate $0.05$ and $0.02$ (that is, $n=16000$ and $n=40000$), 
respectively.

As we observe from their simulated performance, reported in Fig. \ref{fig:OtherRates},
by using the SPA, these codes are able to reach very low error rates
for a percentage of bit errors around $30$\% and even more.
Also in this case, the performance improvement due to the SPA with
respect to the theoretical performance referred to Gallager's A algorithm
is evident.

These results confirm that LDPC codes are well suited for the
application in the considered context, in which a high correction
capability is needed.
Furthermore, we can observe that, in this study, we have limited ourselves to
consider almost regular codes, in order to keep their implementation
complexity low. The adoption of irregular
LDPC codes can result in a further performance improvement.
%So, the use of optimized degree distributions for irregular LDPC
%codes to be used in this context is an interesting option for future
%works on this subject.

\section{Entropy analysis}
\label{sec:EntropyAnalysys}
 
In this section, we discuss the use of fuzzy hashing for iris recognition and we study how
the adoption of syndromes affects some statistical properties of the biometric data.
As a feature extractor, we use the algorithm described in \cite{Masek2003} and available
in \cite{Masek2003web}, together with its associated matching algorithm.
In our simulations, we refer to the iris pattern database known as CASIA V.1,
provided by the Institute of Automation of the Chinese Academy of Sciences \cite{CASIAweb}.
 
%In the matching phase we used the algorithm proposed in the same package (http://www.csse.uwa.edu.au/~pk/studentprojects/libor/sourcecode.html) to evaluate the (average) number of errors we need to correct in two differents templates describing the same iris in order to have a positive match

Since the bits in iris templates are mutually dependent \cite{Daugman2004}, we should compute the entropy 
of a source with memory, but this is computationally unfeasible for the sizes we are dealing with. So, according
to \cite{Daugman2004, Cover1991}, we evaluate the discrimination entropy over
both the sets of iris templates and of their fuzzy hashes.
For this purpose, we first compute the distribution of the normalized Hamming distances 
between all the couples of patterns within the set (of images of the same iris or of images 
of different irises). Then, we compute the mean $\mu$ and the standard deviation $\sigma$ of
the normalized Hamming distance distribution.
Finally, the discrimination entropy (also known as ``Degrees of Freedom'' or DOF) is obtained as 
$\mathrm{DOF} = \mu(1-\mu)/\sigma^2$.

Applying fuzzy hashing to an iris recognition framework is not straightforward, due to
the high variability in the iris acquisition phase. In fact, we must try to avoid all the differences
given not only by the measure variability (i.e., scale and rotation), but also by the eye variability,
that can significantly change the amount of visible iris and its shape.

The standard way to take these issues into account is to compute a mask describing which bits in the iris template are
free from such occlusions.
The masks, in general, have a different number of set bits for each iris reading, resulting into information patterns having different 
lengths, both in the case they describe different irises and different readings of the same iris.
This is not a problem in the case of the standard matching algorithms, since we can take, as inputs for
the matching phase, the templates and masks of both the stored iris and the one we want to check.
Then, we can just compute the union of the two masks and obtain the number
of different bits between the two templates, limited to the region excluded from the union of the masks.

Instead, when we use syndromes, we cannot access the reference template in clear; so, we must cope with different lengths
of the information patterns. 
One way is to treat the matching channel as an error-and-erasure channel \cite{Bringer2007}, where erasures are given by the masks.
However, in \cite{Bringer2007} the authors use a different algorithm, while, in our case, the large number of bits erased by the masks makes this approach unusable.
In order to obtain a fixed length of the information patterns, we compute, for each template bit position $i$, 
the probability $m(i)$ that such bit is not erased by a mask.
Then, we compute a pseudomask selecting the bit positions corresponding to a value of $m(i)$ lower than
a threshold: $m(i) \le m_{th}$. In our case, we fix $m_{th}=2.4\%$.

We are aware that, with this approach, we may neglect some bits that were not erased by their associated masks,
but we have verified that this has a very limited effect for the considered algorithm.
In fact, using all the selected bits in each template, we obtain, between two different readings
of the same iris, an average Hamming distance of $28.24\%$, a standard deviation $\sigma=0.0435$ and a discrimination entropy
equal to $107$ bits. Instead, using only the bits selected by the pseudomask, we obtain an average Hamming distance of $26.2\%$,
a standard deviation $\sigma=0.0486$ and a discrimination entropy equal to $81$ bit.
The explanation for this moderate variation, in terms of discrimination entropy performance, is that the feature extraction
algorithm does not compute the masks in the best possible way (for example the two eyelids are approximated with straight lines
and not with curves); so, when we consider all the bits in each template, we introduce some errors that afflict the result,
that is, we take some bits into account that we should actually erase.

We model the channel as BSC both in the case of intra-class and inter-class comparisons.
The use of a BSC model is further justified by the fact that, by exploiting the considered feature extraction
algorithm and pseudomask, we have an experimental transition probability $p_{1\rightarrow0}=0.257$,
$p_{0\rightarrow1}=0.268$ for intra-class comparison and $p_{1\rightarrow0}=0.480$,
$p_{0\rightarrow1}=0.501$ for the inter-class case, thus confirming the (almost) symmetric nature
of the channel.

\begin{figure}[htb]
\begin{centering}
\includegraphics[width=60mm,keepaspectratio]{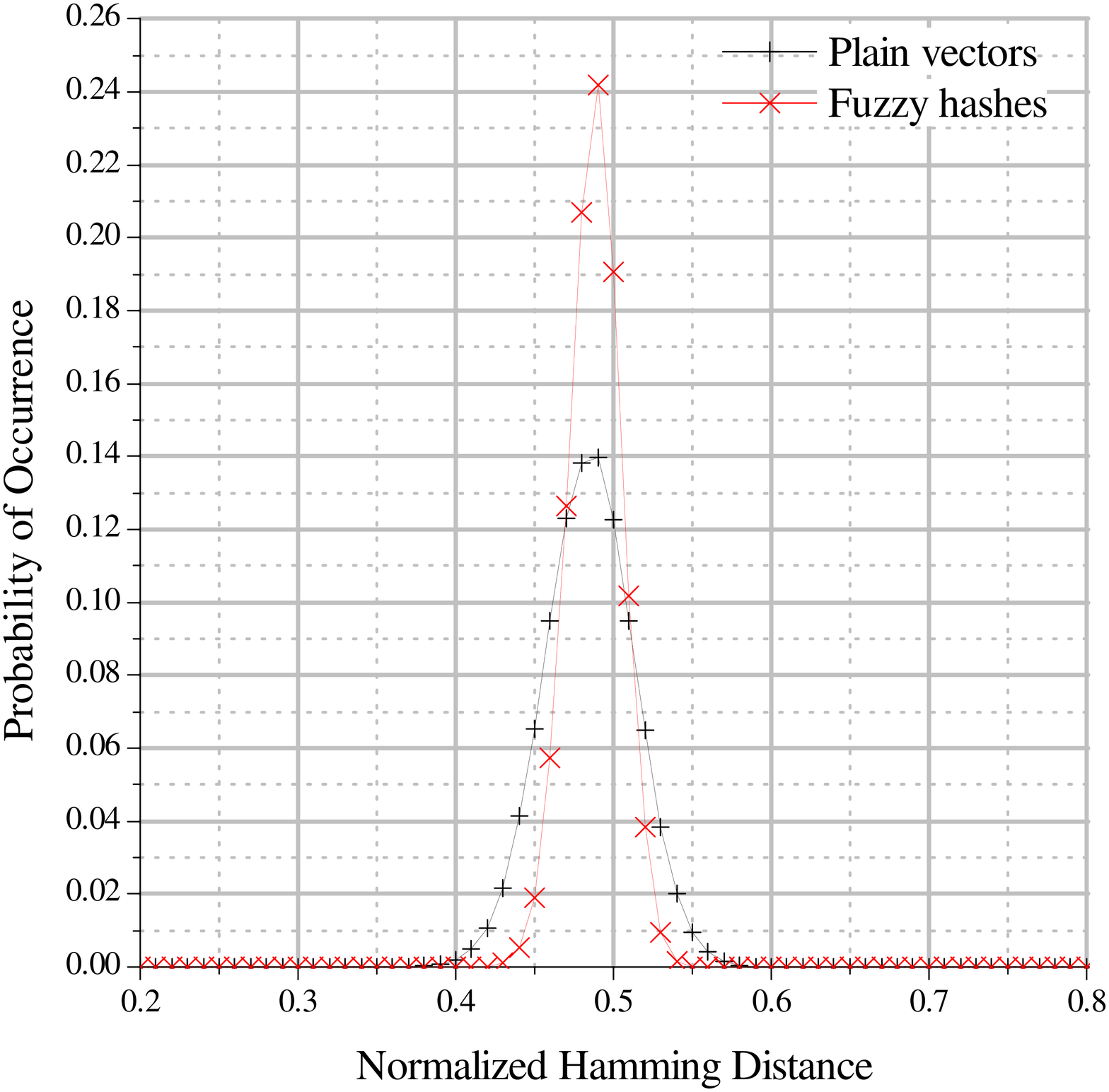}
\caption{Inter-class analysis of the Hamming distance for the considered set of iris templates with and without fuzzy hashing. 
\label{fig:FigHammingDistance}}
\par\end{centering}
\end{figure}

In order to show that the use of fuzzy hashing is able to increase the discrimination entropy, that is to provide a better 
protection against information leakage, we estimate the DOF on the set of plain templates, before and after the application
of fuzzy hashing. The latter is performed through the LDPC code having $n=9600$, $R=0.1$ and $d_v = 5$, described in the
previous section.
We compute the normalized Hamming distance between each pair of templates created from different irises and then
estimate its probability density function. The results are reported in Fig. \ref{fig:FigHammingDistance}.

The set of plain template vectors has $\mu = 0.4897$, $\sigma = 0.0281$, hence $\mathrm{DOF} = 316.5$. After performing
fuzzy hashing, the values become $\mu = 0.4932$, $\sigma = 0.0166$, $\mathrm{DOF} = 907.1$, thus confirming the positive
effect of fuzzy hashing.

\vspace{1em}

%\section{Concluding remarks}

%\section*{Acknowledgment}
%The Research was supported in part by the Swiss National Science
%Foundation under grant No. 132256. 

%\IEEEtriggeratref{26}

% ******************************************************************

\end{document}